\documentclass[conference]{IEEEtran}
\usepackage{cite,graphicx,amsmath,booktabs}
\usepackage{tikz}
\usepackage{url}
\usepackage{array}
\usepackage[colorlinks=true,linkcolor=blue,citecolor=blue,urlcolor=blue]{hyperref}
\usetikzlibrary{arrows.meta,positioning,shapes.geometric,fit}

\begin{document}

\title{GNSS Spoofing Detection in TDD Networks: A 3GPP Standards-Based Security Framework}

\author{\IEEEauthorblockN{Ravi Kant Sharma, John Owens, Kevin Kiernan}
\IEEEauthorblockA{Ericsson\\
\{ravi.kant.sharma, john.owens, kevin.kiernan\}@ericsson.com}}

\maketitle

\begin{abstract}
Time Division Duplex (TDD) mobile networks require synchronization accuracy of $\pm$1.5\,$\mu$s (3GPP TS~38.104), with GNSS-disciplined grandmaster clocks as the predominant timing source. GNSS spoofing---now a documented operational threat---can corrupt timing across all downstream base stations, yet neither the 3GPP management framework (SA5) nor the security framework (SA3) provides standardized mechanisms to detect or report such attacks. This paper proposes a detection and monitoring framework operating within existing 3GPP management structures. The framework introduces GNSS timing alarms and performance counters aligned with TS~28.111 and TS~28.552, a topology-aware correlation mechanism that classifies anomalies by grouping gNB-DUs by serving grandmaster, and a security event bridging fault management with SECHAND incident handling (TR~33.894). Monte Carlo simulation demonstrates detection probability exceeding 95\% for drift rates above 0.5\,ns/s with false positive rates below 1\% under well-provisioned PTP network conditions ($\sigma_{ptp} \leq 5$\,ns). The framework requires no new interfaces, is generation-agnostic, and is validated through scenario analysis distinguishing spoofing from signal loss, equipment faults, and maintenance transients.
\end{abstract}

\begin{IEEEkeywords}
GNSS spoofing, TDD synchronization, 3GPP management, timing security, zero trust, anti-spoofing, 5G/6G network resilience
\end{IEEEkeywords}

\section{Introduction}
\label{sec:introduction}

Fifth-generation (5G) wireless networks and their successors operating in Time Division Duplex (TDD) mode require precise time and phase synchronization between base stations. The 3GPP specification TS~38.104~\cite{ts38104} mandates a timing accuracy of $\pm$1.5\,$\mu$s at the air interface for TDD operation. Violation of this requirement results in inter-cell interference, degraded throughput, and potential service outages. As TDD dominates current 5G deployments---particularly in the mid-band spectrum (3.3--4.2\,GHz and 4.4--5.0\,GHz)---and is expected to remain the primary duplexing mode in 6G, timing accuracy is not merely a quality metric but a fundamental operational requirement that will persist across network generations.

In many mobile network deployments, the primary timing source is the Global Navigation Satellite System (GNSS). In a typical deployment, GNSS receivers embedded in grandmaster clocks derive Coordinated Universal Time (UTC) from satellite signals. This timing is distributed through the transport network via the Precision Time Protocol (PTP, IEEE~1588) following ITU-T telecom PTP profiles~\cite{g8275_1} to boundary clocks and ultimately to the gNB Distributed Units (gNB-DUs)\footnote{Throughout this paper, we use gNB-DU (Distributed Unit) as PTP timing terminates at the DU in the 3GPP split architecture. Where the specific split is not relevant, gNB is used as shorthand.}. The resulting synchronization chain follows the path: GNSS satellite $\rightarrow$ grandmaster clock $\rightarrow$ boundary clocks $\rightarrow$ gNB-DU.

In deployments where a GNSS-disciplined grandmaster is the primary timing source, this architecture introduces a single point of vulnerability at the GNSS interface. GNSS spoofing---the transmission of counterfeit satellite signals that cause receivers to compute incorrect position or time---has transitioned from a theoretical risk to a documented operational threat. In the Baltic region during 2024, aviation authorities recorded 985 GPS disruption events over a two-month period~\cite{baltic2024}. In June 2025, 13~EU Member States formally requested the European Commission to take immediate action against GNSS interference; EASA and EUROCONTROL subsequently published a joint Action Plan in March 2026 establishing coordinated detection, monitoring, and mitigation measures for aviation~\cite{easa2026}. The GNSS signals affected in these incidents are the same signals used by mobile network grandmaster clocks. While aviation has initiated institutional-level coordinated responses, the telecommunications sector---equally dependent on GNSS timing---currently lacks an equivalent common standardized spoofing-specific reporting framework.

The consequences of successful GNSS spoofing against a TDD network are severe. A spoofed grandmaster clock distributes corrupted timing to all downstream gNB-DUs. If the induced timing error exceeds the $\pm$1.5\,$\mu$s budget, TDD guard periods are violated, causing uplink-downlink interference across the affected cells. Unlike a sudden GNSS signal loss---which triggers holdover mode and is typically detectable---sophisticated spoofing attacks introduce gradual timing drift that evades simple threshold-based detection. Prior research has demonstrated that mobile network timing receivers are vulnerable to such spoofing attacks, with the potential to disrupt TDD operation across multiple base stations~\cite{ion_gnss_spoofing}.

The emerging zero-trust security paradigm further motivates this work: GNSS timing is currently implicitly trusted throughout the synchronization chain, with no verification or continuous monitoring at the management layer. Additionally, evolving regulatory frameworks---including the EU Cyber Resilience Act (CRA), the NIS2 Directive, and equivalent national regulations---increasingly require operators and equipment manufacturers to demonstrate security event detection and reporting capabilities across all attack surfaces, including infrastructure dependencies such as timing.

Despite the severity of this threat, a significant gap exists in the 3GPP standards landscape. The SA5 management framework, which defines alarm management (TS~28.111~\cite{ts28111}), performance measurement (TS~28.552~\cite{ts28552}), and the Network Resource Model (TS~28.541~\cite{ts28541}), currently contains no provisions for GNSS timing integrity monitoring. There are no alarm types for timing anomalies, no performance counters for GNSS signal quality, and no managed object representing the grandmaster clock. Similarly, the SA3 security framework, including the Security Handling (SECHAND) procedures defined in TR~33.894~\cite{tr33894}, focuses on Service-Based Architecture (SBA) security events and does not address physical-layer timing attacks.

This paper addresses this gap by proposing extensions to the 3GPP management and security framework that integrate GNSS spoofing detection into the standard network management infrastructure. While individual vendors and operators may deploy proprietary timing-security mechanisms, the absence of common spoofing-specific 3GPP SA5/SA3 artifacts prevents interoperable reporting and cross-vendor correlation. The extensions presented here are designed to operate within existing 3GPP specification structures with minimal disruption, leveraging established information models and service definitions. Our contributions are threefold:

\begin{enumerate}
    \item New alarm definitions and performance management (PM) counters that make GNSS timing anomalies visible to the 3GPP management system, enabling operators to detect and respond to spoofing events using existing management interfaces.
    \item A network-wide timing drift correlation mechanism that classifies anomalies by analyzing PM data across multiple gNB-DUs grouped by their serving grandmaster clock, distinguishing spoofing from equipment faults and signal loss.
    \item The definition of GNSS spoofing as a SECHAND security event, bridging the gap between timing fault management and security incident handling, enabling correlation with other attack vectors.
\end{enumerate}

The paper presents a standards-aligned architectural framework validated through scenario-based analysis and Monte Carlo simulation of detection performance under realistic noise conditions.

The remainder of this paper is organized as follows. Section~\ref{sec:background} provides background on the 3GPP management and security frameworks and reviews related work. Section~\ref{sec:problem} defines the specific gaps in current standards. Section~\ref{sec:framework} presents our proposed framework extensions. Section~\ref{sec:detection} describes the detection architecture. Section~\ref{sec:examples} illustrates network-wide detection through detailed examples. Section~\ref{sec:evaluation} presents simulation-based evaluation of detection performance. Section~\ref{sec:discussion} discusses deployment considerations and limitations. Section~\ref{sec:conclusion} concludes.

\section{Background and Related Work}
\label{sec:background}

\subsection{3GPP SA5 Management Framework}

The 3GPP SA5 working group defines the management architecture for mobile networks through several specifications relevant to this work. TS~28.532~\cite{ts28532} specifies generic management services (provisioning, fault supervision, performance assurance) operating over RESTful interfaces in a producer-consumer model. TS~28.111~\cite{ts28111} defines fault management following the ITU-T X.733 alarm model, characterizing alarms by \texttt{objectClass}, \texttt{alarmType}, \texttt{probableCause}, and \texttt{perceivedSeverity}---but defines no spoofing-specific GNSS timing alarms. TS~28.552~\cite{ts28552} defines performance measurements organized into measurement families collected at configurable granularity periods (typically 15 minutes)---but contains no measurement family for GNSS timing or synchronization quality. TS~28.541~\cite{ts28541} defines the Network Resource Model (NRM) including \texttt{GnbDuFunction} for the gNB Distributed Unit---but does not model the grandmaster clock or the synchronization chain.

\subsection{3GPP SA3 Security Framework}

TS~33.501~\cite{ts33501} defines the 5G security architecture covering authentication, key management, and secure communication, but focuses on signaling and user planes rather than synchronization. TR~33.894~\cite{tr33894} introduces the Security Handling (SECHAND) framework for detecting, reporting, and responding to security events. Current SECHAND scope focuses on SBA security events (unauthorized access, message integrity violations, anomalous signaling); physical-layer timing attacks including GNSS spoofing are outside its defined event types.

\subsection{Existing GNSS Anti-Spoofing Solutions}

Several approaches to GNSS anti-spoofing detection exist, but the representative solutions reviewed below do not define integration with 3GPP SA5 fault/performance management or SA3 security-event handling.

Dimetor~\cite{dimetor2026} proposes detecting GNSS anomalies by correlating cellular network observations with expected GNSS behavior. While this leverages telecom infrastructure for detection, it operates as a standalone system with no integration into 3GPP alarm management, PM counters, or security event handling. The detection results are not exposed through standard management interfaces. The distinction from the present work is that our correlation is organized around 3GPP-managed timing domains---specifically the association between a GNSS-disciplined grandmaster and its downstream \texttt{GnbDuFunction} instances---and is exposed through proposed SA5/SA3-aligned management artifacts.

GPSPatron's GP-Probe~\cite{gpspatron} is a dedicated hardware device that monitors GNSS signal characteristics for spoofing indicators including signal power anomalies, carrier-to-noise ratio deviations, and satellite geometry inconsistencies. As standalone hardware, it provides local detection capability but cannot correlate across network-wide timing sources or report through 3GPP management interfaces.

Honeywell~\cite{honeywell2024} proposes comparing GNSS-derived timing against a chip-scale atomic clock (CSAC) reference. Deviations beyond a threshold indicate potential spoofing. While effective for individual receiver protection, this approach has no mechanism for network-wide correlation or integration with telecom management systems.

\subsection{Standards and Industry Initiatives}

IEEE P1952~\cite{ieeep1952} is a standard under development for resilient Position, Navigation, and Timing (PNT) user equipment. It defines requirements for PNT receivers to detect and mitigate spoofing and jamming. However, P1952 focuses on the receiver equipment itself rather than the management systems that oversee network deployments.

The Alliance for Telecommunications Industry Solutions (ATIS) held dedicated sessions on GNSS spoofing in telecommunications at the Workshop on Synchronization in Telecommunication Systems (WSTS) in 2026~\cite{atiswsts2026}. ATIS SYNC has also published a technical report on resilient timing architecture for mobile networks~\cite{atis_resilient_timing}, addressing timing redundancy at the transport level. These sessions and publications highlighted the growing industry recognition of timing vulnerability but did not propose specific 3GPP management framework extensions.

Prior work on mobile network synchronization~\cite{ericssonreview} addresses timing resilience through transport network redundancy---deploying multiple PTP paths, using assisted partial timing support, and implementing boundary clock selection algorithms. While these measures improve robustness, they do not provide management-layer visibility into GNSS spoofing events and cannot detect attacks that corrupt the primary timing source before distribution.

\subsection{Identified Gap}

None of the existing solutions---whether academic, commercial, or standards-based---integrates GNSS spoofing detection with the 3GPP management framework. This represents a critical gap: the timing source that underpins all TDD operation is not visible through the standardized management and security systems designed to protect the network. Specifically, no prior work addresses:
\begin{itemize}
    \item Alarm definitions for GNSS timing anomalies within TS~28.111
    \item Performance measurement counters for GNSS timing security within TS~28.552
    \item A managed object class for the grandmaster clock within the NRM (TS~28.541)
    \item GNSS spoofing as a security event within SECHAND (TR~33.894)
    \item Network-wide correlation of timing drift across gNBs grouped by grandmaster
\end{itemize}

This gap has system-level operational significance: timing-source manipulation can affect multiple downstream radio nodes while remaining outside spoofing-specific standardized management visibility. As GNSS spoofing tools become commercially available and state-sponsored interference incidents multiply, this motivates the introduction of timing security monitoring within the 3GPP management and security framework. The framework proposed in this paper addresses this gap.

\section{Problem Statement}
\label{sec:problem}

Based on the analysis above, we identify five specific gaps in the current 3GPP standards that collectively leave GNSS spoofing invisible to network management. The threat model assumed throughout this paper is a terrestrial GNSS spoofer operating a single-antenna transmitter in proximity to one or more grandmaster clock GNSS antennas. The attacker can broadcast counterfeit GNSS signals that override authentic signals at the target receiver, but does not have access to the PTP distribution network, the management channel, or the gNB-DU equipment. This corresponds to the most commonly documented real-world GNSS interference scenario and excludes insider threats or attacks on the transport network itself.

\begin{enumerate}
    \item \textbf{No spoofing alarm}: TS~28.111 defines no alarm type or probable cause for GNSS timing anomalies. Grandmaster spoofing events cannot be reported through standardized fault management.
    \item \textbf{No timing PM counters}: TS~28.552 contains no measurement family for GNSS signal quality, spoofing detection events, holdover activations, or drift magnitude.
    \item \textbf{No grandmaster IOC}: TS~28.541 does not model the grandmaster clock or the synchronization topology, preventing the management system from associating gNB-DUs with their timing source.
    \item \textbf{No SECHAND event}: TR~33.894 does not define GNSS spoofing as a security event, preventing correlation with other attack vectors or invocation of security response procedures.
    \item \textbf{No network-wide correlation}: No mechanism exists to correlate timing observations across gNB-DUs grouped by grandmaster, which is essential to distinguish grandmaster-level spoofing from individual equipment faults.
\end{enumerate}

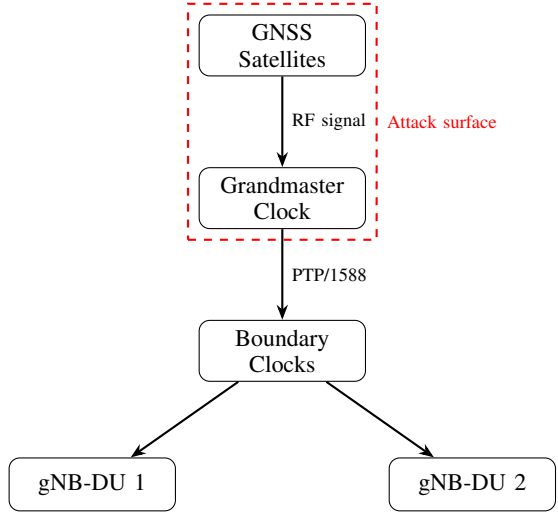
\begin{figure}[t]
\centering
\begin{tikzpicture}[
    node distance=1.2cm,
    every node/.style={font=\small},
    box/.style={draw, rounded corners, minimum width=2.2cm, minimum height=0.7cm, align=center},
    arrow/.style={-{Stealth[length=2mm]}, thick}
]
\node[box] (gnss) {GNSS\\Satellites};
\node[box, below=of gnss] (gm) {Grandmaster\\Clock};
\node[box, below=of gm] (bc) {Boundary\\Clocks};
\node[box, below left=1cm and 0.3cm of bc] (du1) {gNB-DU 1};
\node[box, below right=1cm and 0.3cm of bc] (du2) {gNB-DU 2};

\draw[arrow] (gnss) -- node[right, font=\scriptsize] {RF signal} (gm);
\draw[arrow] (gm) -- node[right, font=\scriptsize] {PTP/1588} (bc);
\draw[arrow] (bc) -- (du1);
\draw[arrow] (bc) -- (du2);

\node[draw, dashed, red, thick, fit=(gnss)(gm), inner sep=4pt, label={[red, font=\scriptsize]right:Attack surface}] {};
\end{tikzpicture}
\caption{TDD synchronization chain showing the GNSS-to-gNB-DU timing path. The attack surface at the GNSS-to-grandmaster interface is currently invisible to the 3GPP management framework.}
\label{fig:sync_chain}
\end{figure}

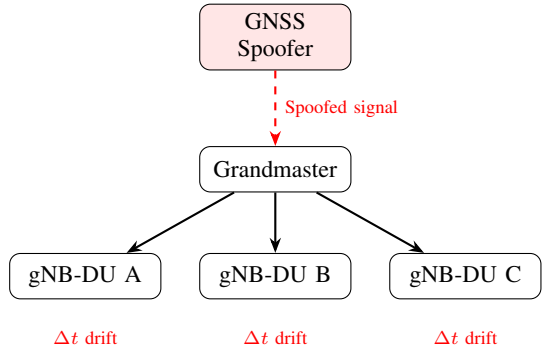
\begin{figure}[t]
\centering
\begin{tikzpicture}[
    node distance=1.0cm,
    every node/.style={font=\small},
    box/.style={draw, rounded corners, minimum width=2.0cm, minimum height=0.6cm, align=center},
    threat/.style={draw, rounded corners, fill=red!10, minimum width=2.0cm, minimum height=0.6cm, align=center},
    arrow/.style={-{Stealth[length=2mm]}, thick}
]
\node[threat] (spoofer) {GNSS\\Spoofer};
\node[box, below=of spoofer] (gm) {Grandmaster};
\node[box, below left=0.8cm and 0.5cm of gm] (du1) {gNB-DU A};
\node[box, below=0.8cm of gm] (du2) {gNB-DU B};
\node[box, below right=0.8cm and 0.5cm of gm] (du3) {gNB-DU C};

\draw[arrow, red, dashed] (spoofer) -- node[right, font=\scriptsize] {Spoofed signal} (gm);
\draw[arrow] (gm) -- (du1);
\draw[arrow] (gm) -- (du2);
\draw[arrow] (gm) -- (du3);

\node[below=0.3cm of du1, font=\scriptsize, red] {$\Delta t$ drift};
\node[below=0.3cm of du2, font=\scriptsize, red] {$\Delta t$ drift};
\node[below=0.3cm of du3, font=\scriptsize, red] {$\Delta t$ drift};
\end{tikzpicture}
\caption{GNSS spoofing attack vector: a single spoofed grandmaster propagates timing errors to all downstream gNB-DUs, creating correlated drift observable at the network level.}
\label{fig:attack_vector}
\end{figure}

Figures~\ref{fig:sync_chain} and~\ref{fig:attack_vector} illustrate the synchronization chain and the attack propagation model.

\section{Proposed Framework}
\label{sec:framework}

We present extensions to three areas of the 3GPP framework: fault management (alarms), performance management (PM counters), and security event handling (SECHAND). Additionally, we propose a network-wide correlation mechanism that leverages these extensions. The framework is designed to operate within existing 3GPP specification structures, using established information models and service definitions without requiring new interfaces or protocols.

\subsection{New Alarm Definitions}
\label{sec:alarms}

We propose two alarm types, modeled on the existing alarm structure defined by ITU-T X.733 and the 3GPP profile in TS~28.111. Both alarms are raised by a \texttt{GnssGrandmaster} managed object class that we introduce to represent the grandmaster clock in the synchronization chain.

Table~\ref{tab:alarms} presents the proposed alarm definitions. The warning-level alarm is raised when timing deviation exceeds a configurable threshold, indicating a potential anomaly that may be spoofing or may resolve naturally. The critical alarm is raised when the detection system has high confidence that spoofing is occurring, based on multiple indicators or sustained deviation.

The warning alarm uses \texttt{qualityOfServiceAlarm} as the alarm type and \texttt{gnssTimingDeviation} as the probable cause, reflecting that at this stage the anomaly may have multiple explanations. The critical alarm uses \texttt{securityServiceOrMechanismViolation} as the alarm type and \texttt{gnssSpoofingDetected} as the probable cause, explicitly classifying the event as a security violation.

\begin{table*}[t]
\centering
\caption{GNSS Spoofing Alarm Definitions (TS~28.111 Structure)}
\label{tab:alarms}
\begin{tabular}{@{}lll@{}}
\toprule
\textbf{Attribute} & \textbf{Warning Alarm} & \textbf{Critical Alarm} \\
\midrule
objectClass & GnssGrandmaster & GnssGrandmaster \\
alarmType & qualityOfServiceAlarm & securityServiceOrMechanismViolation \\
probableCause & gnssTimingDeviation & gnssSpoofingDetected \\
perceivedSeverity & warning & critical \\
specificProblem & Timing drift exceeds threshold & GNSS spoofing attack confirmed \\
proposedRepairAction & Monitor; verify GNSS signal & Activate holdover; isolate timing source \\
additionalText & Drift magnitude, duration, & Detection method, confidence, \\
 & affected downstream gNBs & affected downstream gNBs \\
\bottomrule
\end{tabular}
\end{table*}

As an illustration, the escalation from warning to critical could follow a process such as the following; the exact thresholds, logic, and escalation criteria are implementation-specific. A warning alarm is raised when the timing drift magnitude exceeds a configurable warning threshold $\theta_w$ (e.g., 50\,ns) or when GNSS signal integrity indicators exceed anomaly thresholds. The alarm escalates to critical when: (a) drift exceeds a critical threshold $\theta_c$ (e.g., 200\,ns) with continued signal lock, (b) multiple independent detection methods concur, or (c) network-wide correlation confirms correlated drift across the grandmaster group.

The \texttt{GnssGrandmaster} Information Object Class (IOC) is defined with attributes consistent with the NRM modeling conventions of TS~28.541. Key attributes include:

\begin{itemize}
    \item \texttt{gnssGrandmasterId}: Unique identifier
    \item \texttt{oscillatorType}: Enumeration (CSAC, rubidium, OCXO, TCXO)
    \item \texttt{holdoverCapability}: Maximum holdover duration at specified accuracy (seconds)
    \item \texttt{servedGnbDuList}: List of \texttt{GnbDuFunction} instances receiving timing from this grandmaster
    \item \texttt{gnssConstellations}: Supported GNSS constellations (GPS, Galileo, GLONASS, BeiDou)
    \item \texttt{currentTimingState}: Enumeration (locked, holdover, freerun, spoofingDetected)
    \item \texttt{lastLockTime}: Timestamp of last confirmed GNSS lock
    \item \texttt{currentDriftNs}: Current measured timing drift in nanoseconds
    \item \texttt{siteLocation}: Physical site identifier (enables site-level correlation)
\end{itemize}

The \texttt{servedGnbDuList} attribute establishes the grandmaster-to-gNB relationship required for network-wide correlation (Section~\ref{sec:correlation}). The management system uses this attribute to group gNBs by grandmaster source. The relationship is modeled as a reference association rather than containment, as the grandmaster is a Physical Network Function (PNF) external to the gNB managed object hierarchy.

\subsection{New Performance Management Counters}
\label{sec:pm}

We propose a new measurement family ``GNSS Timing Security'' following the conventions of TS~28.552. Table~\ref{tab:pm} presents the counter definitions, their associated managed object classes, measurement types, and descriptions.

\begin{table*}[t]
\centering
\caption{GNSS Timing Security PM Counters (TS~28.552 Structure)}
\label{tab:pm}
\begin{tabular}{@{}llll@{}}
\toprule
\textbf{Counter Name} & \textbf{Object Class} & \textbf{Type} & \textbf{Description} \\
\midrule
GnssSpoofing.WarningCount & GnssGrandmaster & CC & Cumulative spoofing warning alarms raised \\
GnssSpoofing.DetectedCount & GnssGrandmaster & CC & Cumulative confirmed spoofing detections \\
GnssSpoofing.WarningResolvedCount & GnssGrandmaster & CC & Warnings resolved without escalation \\
GnssHoldover.ActivationCount & GnssGrandmaster & CC & Holdover mode activations \\
GnssHoldover.Duration & GnssGrandmaster & DER & Cumulative holdover duration (seconds) \\
GnssSignal.AnomalyCount & GnssGrandmaster & CC & GNSS signal integrity anomalies detected \\
GnssTiming.MaxDriftNs & GnbDuFunction & GAUGE & Maximum timing drift in nanoseconds \\
\bottomrule
\end{tabular}
\end{table*}

Counter types follow TS~28.552 conventions: CC denotes a cumulative counter that increments monotonically within a granularity period, DER denotes a derived measurement computed from other values, and GAUGE denotes an instantaneous value sampled at the reporting boundary.

The \texttt{GnssTiming.MaxDriftNs} counter is particularly significant for network-wide correlation. Collected from each \texttt{GnbDuFunction}, it reports the maximum absolute estimated timing offset observed during the granularity period, relative to a configured timing reference. The formal measurement definition is as follows:

\begin{itemize}
    \item \textbf{Measured quantity}: Maximum absolute timing offset (nanoseconds) between the gNB-DU's received PTP timing and an implementation-specific reference during the granularity period.
    \item \textbf{Measurement source}: Implementation-dependent; may be derived from PTP servo offset statistics, local oscillator comparison against the PTP-distributed reference, comparison against a backup timing source, or boundary-clock telemetry.
    \item \textbf{Collection method}: Sampled maximum over the granularity period (not instantaneous gauge).
    \item \textbf{Unit}: Nanoseconds (unsigned).
    \item \textbf{Granularity period}: Configurable; default 15 minutes, escalatable to 15 seconds.
    \item \textbf{Holdover behavior}: During holdover, the counter reports drift relative to the holdover oscillator's expected trajectory.
    \item \textbf{Unavailability}: If no drift estimate is available, the counter is omitted from the PM report for that period.
\end{itemize}

The specific measurement source and its associated uncertainty are implementation-dependent; the framework requires only that a comparable drift estimate is available across gNB-DUs in a grandmaster group. When multiple gNBs served by the same grandmaster report similar drift values simultaneously, this indicates a grandmaster-level anomaly rather than an individual equipment fault.

The measurement granularity period is configurable per TS~28.552 provisions. For timing security, shorter periods (15 seconds to 1 minute) may be appropriate to enable rapid detection of fast-moving spoofing attacks, while standard 15-minute periods suffice for trend analysis and reporting. This introduces an inherent detection latency equal to the granularity period; at 15-second granularity, the minimum time from attack onset to first observable anomaly in PM data is 15 seconds. For slow-drift attacks near the warning threshold (drift rate $\approx$0.5\,ns/s), accumulation to $\theta_w = 50$\,ns requires approximately 100 seconds; reaching the $\pm$1.5\,$\mu$s TDD budget at this rate would take approximately 50 minutes, providing substantial operational margin for detection and response even at standard reporting granularity.

The PM reporting volume at shorter granularity periods requires consideration. At 15-second granularity, each grandmaster generates 5,760 PM reports per day for the six grandmaster-level counters. For a network with 100 grandmasters, this yields approximately 576,000 daily reports---an order of magnitude increase over standard 15-minute reporting. One possible approach to address this is a tiered collection strategy: standard 15-minute reporting for baseline trend analysis under normal conditions, with escalation to shorter granularity triggered by alarm activation on a specific grandmaster, returning to standard reporting after alarm clearance. The specific escalation triggers and granularity levels are implementation decisions that depend on network scale, management system capacity, and operator requirements.

\subsection{SECHAND Security Event Definition}
\label{sec:sechand}

We define a security event type that bridges timing fault management with security incident handling, following the SECHAND framework structure defined in TR~33.894. An illustrative security event structure includes:

\begin{itemize}
    \item \texttt{securityEventType}: \texttt{gnssSpoofingAttack}
    \item \texttt{sourceObject}: \texttt{GnssGrandmaster} (the grandmaster under attack)
    \item \texttt{severity}: critical
    \item \texttt{affectedObjects}: list of downstream \texttt{GnbDuFunction} instances receiving timing from the compromised grandmaster
    \item \texttt{detectionMethod}: enumeration indicating the detection mechanism (oscillator comparison, signal integrity, network correlation, or combination)
    \item \texttt{confidence}: numerical confidence score (0.0--1.0)
    \item \texttt{attackCharacteristics}: drift rate, drift direction, duration, number of affected gNBs
\end{itemize}

This security event type serves three purposes. First, it enables the security management system to be aware of timing attacks, which are otherwise handled only by fault management. Second, it enables correlation between timing attacks and other security events. A coordinated attack might combine GNSS spoofing with a Distributed Denial of Service (DDoS) attack against the core network; without a SECHAND event for spoofing, the security system cannot correlate these events or identify a coordinated campaign. Third, it could support security response procedures aligned with SECHAND concepts, including notification to security operations, automated mitigation actions, and forensic data collection.

The relationship between fault management alarms and the SECHAND security event is defined as follows: the critical alarm (\texttt{gnssSpoofingDetected}) raised by fault management triggers generation of the SECHAND security event. The warning alarm does not trigger a security event, as it may represent legitimate anomalies. This separation ensures that security response procedures are invoked only when spoofing is confirmed with sufficient confidence.

\subsection{Network-Wide Timing Drift Correlation}
\label{sec:correlation}

The most significant contribution of this work is the network-wide timing drift correlation mechanism. Individual detection at a single grandmaster or gNB has inherent limitations: oscillator comparison depends on local oscillator quality, and signal integrity monitoring can be evaded by sophisticated spoofers. Network-wide correlation adds a dimension that individual detection cannot provide. \textbf{Independence requirement.} An important prerequisite is that the gNB-DU drift estimate must be referenced against something at least partially independent of the compromised grandmaster chain. If all DUs faithfully track a cleanly-spoofed grandmaster via PTP with no independent reference, they report near-zero drift despite accumulating absolute timing error---rendering correlation-based detection ineffective. The mechanism is therefore most effective when DUs derive their drift estimate from: (a)~a local holdover oscillator free-running alongside the PTP-disciplined clock, (b)~a backup timing source such as a secondary GNSS receiver or ePRTC, (c)~boundary-clock telemetry reporting PTP servo statistics (offset, path delay variation), or (d)~cross-comparison between neighboring boundary clocks on different PTP paths. In current deployments, commercial gNB-DU implementations that comply with ITU-T G.8275.1 maintain PTP servo offset statistics that can provide case~(c): the servo continuously measures the offset between its local oscillator and the received PTP timing, and abrupt changes in this offset---even when the servo tracks them---are observable as transient excursions in the servo error signal. For deployments where DUs have no independent timing reference whatsoever, Method~3 correlation must be supplemented by Methods~1 and~2 at the grandmaster level. The framework explicitly accommodates this limitation: the \texttt{GnssTiming.MaxDriftNs} counter definition specifies that the measurement source is implementation-dependent, and the counter is omitted when no drift estimate is available.

\begin{figure}[t]
\centering
\begin{tikzpicture}[
    node distance=0.6cm,
    every node/.style={font=\scriptsize},
    box/.style={draw, rounded corners, minimum width=1.2cm, minimum height=0.5cm, align=center},
    mgmt/.style={draw, rounded corners, fill=blue!10, minimum width=1.8cm, minimum height=0.5cm, align=center},
    arrow/.style={-{Stealth[length=2mm]}, thick}
]
\node[mgmt] (ms) {Management\\System};
\node[box, below left=1.0cm and 0.8cm of ms] (gm1) {GM-A};
\node[box, below right=1.0cm and 0.8cm of ms] (gm2) {GM-B};
\node[box, below left=0.7cm and 0.2cm of gm1] (du1) {DU-A};
\node[box, below=0.7cm of gm1] (du2) {DU-B};
\node[box, below right=0.7cm and 0.2cm of gm1] (du3) {DU-C};
\node[box, below left=0.7cm and 0.15cm of gm2] (du4) {DU-D};
\node[box, below right=0.7cm and 0.15cm of gm2] (du5) {DU-E};

\node[draw, dashed, gray, rounded corners, fit=(gm1)(du1)(du2)(du3), inner sep=3pt, label={[gray, font=\tiny]below:Group A}] {};
\node[draw, dashed, gray, rounded corners, fit=(gm2)(du4)(du5), inner sep=3pt, label={[gray, font=\tiny]below:Group B}] {};

\draw[arrow] (gm1) -- (du1);
\draw[arrow] (gm1) -- (du2);
\draw[arrow] (gm1) -- (du3);
\draw[arrow] (gm2) -- (du4);
\draw[arrow] (gm2) -- (du5);
\draw[arrow, dashed] (gm1) -- node[left, font=\tiny] {alarms/PM} (ms);
\draw[arrow, dashed] (gm2) -- node[right, font=\tiny] {alarms/PM} (ms);
\draw[arrow, dashed] (du1.north west) to[bend left=20] (ms.south west);
\draw[arrow, dashed] (du5.north east) to[bend right=20] (ms.south east);
\end{tikzpicture}
\caption{System architecture: the management system collects PM data and alarms from grandmasters and gNB-DUs, enabling network-wide correlation of timing drift across grandmaster groups.}
\label{fig:architecture}
\end{figure}
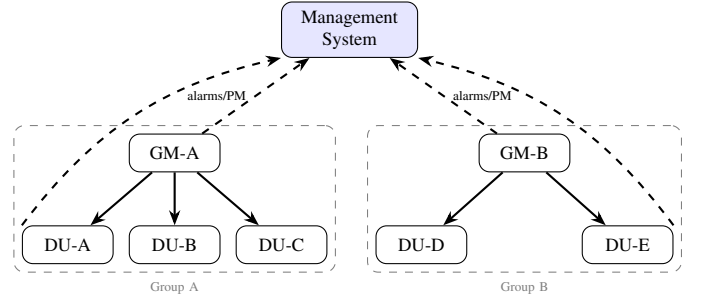

The correlation mechanism operates in the management system, consuming PM data (\texttt{GnssTiming.MaxDriftNs}) from all \texttt{GnbDuFunction} instances and grouping them by their serving grandmaster (Figure~\ref{fig:architecture}). One illustrative approach implements a two-stage decision process:

\textbf{Stage 1: Correlated Drift Detection.} The management system computes pairwise correlation between timing drift time-series across all gNB-DUs in a grandmaster group, using a sliding window of configurable duration at the configured granularity period. Prior to correlation, samples are time-aligned to a common granularity boundary; missing samples are excluded from the pairwise computation. Drift slopes across the group are also compared to avoid classifying unrelated monotonic trends as common-mode drift. As an illustrative example, the correlation threshold $\rho_{threshold}$ may be set at 0.85 for groups of 3 or more gNB-DUs, requiring that the minimum pairwise correlation within the group exceeds this value. For smaller groups (2 gNB-DUs), a higher threshold (e.g., 0.95) compensates for reduced statistical confidence. Additionally, a minimum drift magnitude threshold (e.g., 10\,ns) avoids triggering on measurement noise. Groups with fewer than 2 gNB-DUs cannot use correlation-based detection and must rely solely on Methods~1 and~2 at the grandmaster level. These illustrative thresholds are informed by the following considerations. The drift magnitude threshold should exceed the typical PTP servo noise floor: commercial PTP implementations in well-provisioned networks with full timing support report steady-state servo offset jitter of 3--5\,ns (1$\sigma$), so a threshold at approximately $2\sigma$ of the combined noise distribution (PTP jitter plus oscillator phase noise) yields acceptable false trigger probability. The group correlation threshold balances detection sensitivity against false correlation from environmental effects. The specific threshold values, window durations, and decision logic are implementation and deployment-specific choices that depend on network topology, oscillator characteristics, and operator risk tolerance.

\textbf{Stage 2: Alarm Verification.} The correlated drift detection is verified against the grandmaster alarm state, as illustrated in Figure~\ref{fig:decision_tree}. Three outcomes are possible:
\begin{itemize}
    \item Correlated drift + spoofing alarm $\rightarrow$ confirmed spoofing attack
    \item Correlated drift + signal-loss alarm $\rightarrow$ GNSS outage (holdover)
    \item Correlated drift + no alarm $\rightarrow$ potential undetected attack or transport issue
\end{itemize}

Classification uses three dimensions:
\begin{enumerate}
    \item \textbf{Scope}: Which gNBs are affected? All gNBs under one grandmaster (grandmaster attack), all gNBs at one site (site-level attack), or a single gNB (individual fault).
    \item \textbf{Trend}: How does the drift behave? Monotonically increasing (spoofing), step-change then stable (holdover), or random/oscillating (equipment fault).
    \item \textbf{GM Alarm}: What alarm has the grandmaster raised? Spoofing detected, signal lost, or no alarm.
\end{enumerate}

The proposed framework is deliberately technique-agnostic: it defines \emph{what} data is collected and \emph{how} anomalies are reported through standardized management interfaces, but does not prescribe \emph{which} analytical method an implementation must use for drift detection or classification. Any suitable technique---whether statistical correlation, change-point detection, trend tests, anomaly detection, time-series forecasting, or supervised classification---may be employed by the management system, provided it consumes the defined PM inputs and produces the defined alarm and event outputs. This separation of interface from algorithm enables vendors to innovate on detection quality without requiring specification updates.

The correlation algorithm is an implementation choice; the framework defines only the measurements and alarm/event interfaces. Pearson correlation is used as an illustrative baseline metric rather than a complete standalone detector. Its known limitations in this context include: (a)~sensitivity to outliers from packet-delay variation or measurement artifacts, (b)~inflation of apparent significance due to autocorrelation in timing-drift time series, (c)~risk of false positives from benign common-mode trends such as temperature cycling, PTP path reconfiguration, or coordinated maintenance, and (d)~weak statistical power in small groups (two gNB-DUs). In operational deployments, correlation should be combined with drift magnitude, slope consistency, persistence criteria, grandmaster alarm state, and maintenance-context information. Alternative or complementary metrics---including Spearman rank correlation for nonlinear drift, change-point detection for attack onset, and robust estimators for outlier tolerance---may improve classification accuracy and are compatible with the proposed framework.

\begin{figure}[t]
\centering
\begin{tikzpicture}[
    node distance=0.8cm,
    every node/.style={font=\scriptsize},
    box/.style={draw, rounded corners, minimum width=2.2cm, minimum height=0.5cm, align=center},
    decision/.style={draw, diamond, aspect=2, minimum width=1.2cm, align=center},
    arrow/.style={-{Stealth[length=2mm]}, thick}
]
\node[box] (collect) {Collect PM data\\from all gNB-DUs};
\node[box, below=of collect] (group) {Group by\\grandmaster};
\node[decision, below=0.9cm of group] (corr) {Correlated\\drift?};
\node[decision, below left=1cm and 0.0cm of corr] (alarm) {GM alarm\\type?};
\node[box, below left=0.8cm and 0.3cm of alarm] (spoof) {Confirmed\\Spoofing};
\node[box, below right=0.8cm and 0.3cm of alarm] (outage) {GNSS\\Outage};
\node[box, below right=0.8cm and 0.5cm of corr] (individual) {Individual\\Fault};

\draw[arrow] (collect) -- (group);
\draw[arrow] (group) -- (corr);
\draw[arrow] (corr) -- node[left] {Yes} (alarm);
\draw[arrow] (corr) -- node[right] {No} (individual);
\draw[arrow] (alarm) -- node[left, font=\tiny] {Spoofing} (spoof);
\draw[arrow] (alarm) -- node[right, font=\tiny] {Signal loss} (outage);
\end{tikzpicture}
\caption{Network-wide correlation decision tree: PM data from gNB-DUs is grouped by grandmaster, analyzed for correlated drift, and classified based on the grandmaster alarm state.}
\label{fig:decision_tree}
\end{figure}
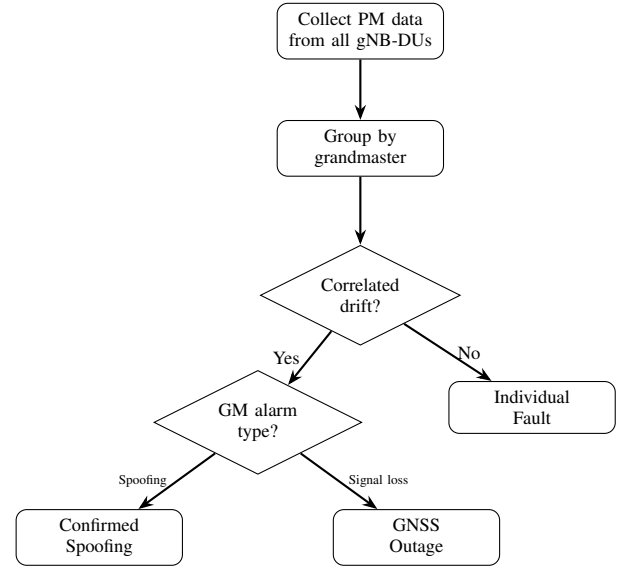

\section{Detection Architecture}
\label{sec:detection}

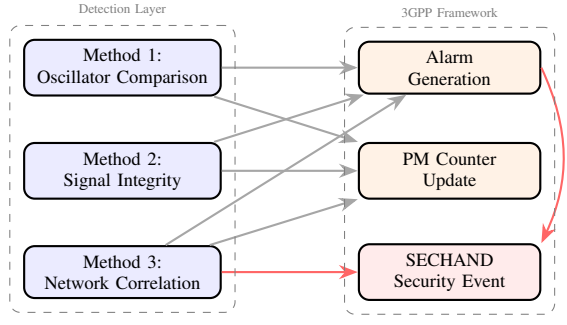
\begin{figure}[t]
\centering
\begin{tikzpicture}[
    node distance=0.6cm and 1.8cm,
    every node/.style={font=\scriptsize},
    detect/.style={draw, rounded corners, fill=blue!8, minimum width=2.6cm, minimum height=0.6cm, align=center, thick},
    output/.style={draw, rounded corners, fill=orange!10, minimum width=2.4cm, minimum height=0.6cm, align=center, thick},
    arrow/.style={-{Stealth[length=2.5mm]}, thick, gray!70},
    critarrow/.style={-{Stealth[length=2.5mm]}, thick, red!60}
]
\node[detect] (m1) {Method 1:\\Oscillator Comparison};
\node[detect, below=of m1] (m2) {Method 2:\\Signal Integrity};
\node[detect, below=of m2] (m3) {Method 3:\\Network Correlation};

\node[output, right=of m1] (alarm) {Alarm\\Generation};
\node[output, right=of m2] (pm) {PM Counter\\Update};
\node[output, fill=red!8, right=of m3] (sec) {SECHAND\\Security Event};

\draw[arrow] (m1) -- (alarm);
\draw[arrow] (m1) -- (pm);
\draw[arrow] (m2) -- (alarm);
\draw[arrow] (m2) -- (pm);
\draw[arrow] (m3) -- (alarm);
\draw[arrow] (m3) -- (pm);
\draw[critarrow] (m3) -- (sec);
\draw[critarrow] (alarm.east) to[bend left=25] (sec.north east);

\node[draw, dashed, gray, rounded corners, fit=(m1)(m2)(m3), inner sep=5pt, label={[gray, font=\tiny]above:Detection Layer}] {};
\node[draw, dashed, gray, rounded corners, fit=(alarm)(pm)(sec), inner sep=5pt, label={[gray, font=\tiny]above:3GPP Framework}] {};
\end{tikzpicture}
\caption{Detection flow: three detection methods feed into the 3GPP framework through alarm generation, PM counter updates, and SECHAND security events. Network-wide correlation (Method~3) can directly trigger security events.}
\label{fig:detection_flow}
\end{figure}

This section describes the detection methods that generate the alarms and PM data defined in Section~\ref{sec:framework}. We emphasize that the detection methods themselves are not the primary contribution of this work; rather, the contribution lies in their structured integration with the 3GPP management and security framework, enabling network-wide correlation. The detection architecture (Figure~\ref{fig:detection_flow}) provides the inputs that the framework extensions consume.

\subsection{Method 1: GNSS vs.\ Independent Oscillator Comparison}

The first detection method compares GNSS-derived time against an independent local oscillator. The grandmaster clock maintains both a GNSS-disciplined oscillator and an independent reference oscillator (rubidium, OCXO, or CSAC). Under normal operation, the time difference $\Delta t$ between the two sources remains within the oscillator's stability specification. An increasing $\Delta t$ beyond expected drift indicates that either the GNSS signal or the local oscillator has deviated.

Since the local oscillator's drift characteristics are known and bounded (e.g., a rubidium oscillator has fractional frequency offset approximately $\pm 1 \times 10^{-11}$), deviations exceeding this bound are attributed to the GNSS source. The detection logic implements a two-threshold scheme:

\begin{equation}
\text{Alarm} = \begin{cases}
\text{warning} & \text{if } |\Delta t| > \theta_w \\
\text{critical} & \text{if } |\Delta t| > \theta_c \text{ and GNSS lock maintained}
\end{cases}
\label{eq:threshold}
\end{equation}

The condition ``GNSS lock maintained'' is critical: if the GNSS receiver reports loss of lock, the deviation is attributed to signal loss rather than spoofing, and holdover mode is activated instead.

The threshold values are illustrative and are derived from the TDD timing budget and oscillator specifications. The critical threshold $\theta_c$ is set relative to the TDD timing error budget: with a total budget of $\pm$1.5\,$\mu$s and the synchronization network allocated a portion of this budget (ITU-T G.8271~\cite{g8271} defines the detailed breakdown), the grandmaster contribution must remain well below the total budget. Setting $\theta_c = 200$\,ns is a conservative choice that leaves substantial operational margin before the TDD alignment budget is exhausted. In practice, thresholds should be calibrated per deployment considering the applicable ITU-T timing specifications~\cite{g8272} (e.g., G.8272 for PRTC/ePRTC), oscillator class, local noise floor, and operator false-positive tolerance. The warning threshold $\theta_w$ is derived from the independent oscillator's fractional frequency offset: for a rubidium oscillator with offset $\pm 1 \times 10^{-11}$ (representing a well-calibrated unit), the expected drift over a 1-hour observation window is approximately 36\,ns. Setting $\theta_w = 50$\,ns (approximately $1.4\times$ the expected oscillator drift over that interval) provides detection sensitivity while avoiding false alarms from normal oscillator behavior. For deployments using lower-quality oscillators (OCXO: fractional frequency offset $\pm 1 \times 10^{-9}$), the observation window must be shortened substantially, as a 1-hour window at this offset accumulates 3.6\,$\mu$s---exceeding the TDD budget itself. OCXO-based detection therefore uses shorter comparison intervals (e.g., 5--10 minutes), and $\theta_w$ and $\theta_c$ are set relative to the timing budget constraint rather than derived solely from oscillator drift. Table~\ref{tab:thresholds} summarizes illustrative thresholds by oscillator class.

\begin{table}[t]
\centering
\caption{Illustrative Detection Thresholds by Oscillator Class}
\label{tab:thresholds}
\begin{tabular}{@{}llll@{}}
\toprule
\textbf{Oscillator} & \textbf{Freq.\ Offset} & $\boldsymbol{\theta_w}$ & $\boldsymbol{\theta_c}$ \\
\midrule
CSAC & $\pm 3 \times 10^{-12}$ & 20\,ns & 200\,ns \\
Rubidium & $\pm 1 \times 10^{-11}$ & 50\,ns & 200\,ns \\
OCXO & $\pm 1 \times 10^{-9}$ & 500\,ns & 1000\,ns \\
\bottomrule
\end{tabular}
\end{table}

\subsection{Method 2: GNSS Signal Integrity Monitoring}

The second detection method analyzes GNSS signal characteristics for spoofing indicators. Multiple signal-level metrics are monitored:

\begin{itemize}
    \item \textbf{Carrier-to-Noise Ratio (C/N$_0$)}: Spoofing signals typically exhibit elevated and uniform C/N$_0$ across satellites, unlike the natural variation in authentic signals caused by atmospheric conditions and satellite elevation angles.
    \item \textbf{Satellite Count and Geometry}: Sudden changes in the number of visible satellites or in the Geometric Dilution of Precision (GDOP) may indicate signal manipulation.
    \item \textbf{Signal Power Distribution}: Authentic GNSS signals arrive at predictable power levels based on satellite elevation. Spoofing transmitters, operating from a single terrestrial location, produce signals with power distributions inconsistent with expected satellite geometry.
    \item \textbf{Navigation Message Consistency}: Cross-checking navigation message content against known ephemeris data and across multiple GNSS constellations (GPS, Galileo, GLONASS, BeiDou) can reveal inconsistencies introduced by spoofing.
\end{itemize}

Each metric contributes to an anomaly score. When the composite anomaly score exceeds a threshold, the \texttt{GnssSignal.AnomalyCount} PM counter is incremented and, depending on severity, a warning alarm is raised.

\subsection{Method 3: Network-Wide Correlation}

The third detection method operates at the management system level rather than at the individual grandmaster. It compares timing drift measurements (\texttt{GnssTiming.MaxDriftNs}) across all gNBs served by the same grandmaster. This method can detect spoofing even when the individual grandmaster's detection mechanisms fail to identify the attack, provided the attack induces observable drift at the gNB level.

The network-wide correlation implements one possible realization of the two-stage decision process described in Section~\ref{sec:correlation}:

\begin{enumerate}
    \item Compute pairwise correlation of drift time-series within each grandmaster group.
    \item If correlation exceeds threshold and drift magnitude is non-trivial, flag the grandmaster group.
    \item Cross-reference with grandmaster alarm state to classify the anomaly.
\end{enumerate}

This method is particularly effective against sophisticated spoofing attacks that evade receiver-level detection by introducing drift below individual detection thresholds but above the noise floor observable through network-wide analysis.

\subsection{Detection Limitations and Adversary Adaptation}

We acknowledge that no detection framework is immune to a sufficiently resourced adversary. Specific limitations include:

\begin{itemize}
    \item \textbf{Method 1}: Effective only for attacks that exceed the independent oscillator's drift rate. A CSAC with fractional frequency offset $\pm 3 \times 10^{-12}$ accumulates approximately 259\,ns/day of free-running time error; rubidium at $\pm 1 \times 10^{-11}$ accumulates approximately 864\,ns/day. Spoofing-induced drift that remains below the oscillator's own drift rate over the observation interval is invisible to oscillator comparison alone.
    \item \textbf{Method 2}: Sophisticated spoofers can increasingly replicate authentic signal characteristics. As spoofing technology advances, signal integrity monitoring must evolve correspondingly. This method provides defense-in-depth but should not be relied upon as the sole detection mechanism.
    \item \textbf{Method 3}: Requires a minimum group size of 2 gNB-DUs per grandmaster; networks with 1:1 grandmaster-to-gNB-DU ratios (common in rural or small-cell deployments) cannot use correlation-based detection and must rely solely on Methods~1 and~2 at the grandmaster level. Additionally, an attacker who introduces decorrelated drift (different rates to different downstream gNB-DUs) could evade group correlation, though this requires the attacker to control the PTP distribution network rather than the GNSS signal alone---a scenario outside the threat model defined in Section~\ref{sec:problem}.
\end{itemize}

The framework's strength lies in combining multiple detection dimensions (oscillator, signal, network) such that evading all simultaneously requires substantially greater adversary capability than evading any single method. This paper presents a management integration framework; the specific detection algorithms and their accuracy are implementation-dependent and will benefit from ongoing refinement as adversary capabilities evolve.

\section{Network-Wide Detection Examples}
\label{sec:examples}

To illustrate the correlation mechanism, we present five scenarios using a reference network topology. The topology comprises two grandmasters: GM-A serves three gNB-DUs (DU-A, DU-B, DU-C), and GM-B serves two gNB-DUs (DU-D, DU-E). PM counters are collected at a 15-second granularity period. The \texttt{GnssTiming.MaxDriftNs} counter at each gNB-DU reports the timing drift relative to the expected reference in nanoseconds.

\subsection{Example A: Slow-Drift Spoofing Attack}

In this scenario, an attacker initiates a gradual spoofing attack against GM-A, introducing a slowly increasing timing offset. The spoofing signal is carefully crafted to avoid triggering the grandmaster's immediate detection threshold by maintaining GNSS lock and keeping the initial drift rate below the warning threshold.

Table~\ref{tab:example_a} shows the timing drift values observed across all gNB-DUs over time. The drift in GM-A's group (DU-A, DU-B, DU-C) is highly correlated and monotonically increasing, while GM-B's group (DU-D, DU-E) remains stable.

\begin{table}[t]
\centering
\caption{Example A: Timing Drift (ns) During Slow-Drift Spoofing of GM-A}
\label{tab:example_a}
\begin{tabular}{@{}lccccc@{}}
\toprule
\textbf{Time} & \textbf{DU-A} & \textbf{DU-B} & \textbf{DU-C} & \textbf{DU-D} & \textbf{DU-E} \\
\midrule
$T+0$ & 0 & +1 & $-1$ & 0 & +1 \\
$T+15$\,s & +9 & +6 & +10 & $-1$ & +2 \\
$T+30$\,s & +19 & +15 & +20 & +1 & $-2$ \\
$T+60$\,s & +39 & +34 & +40 & $-1$ & +1 \\
$T+90$\,s & +58 & +53 & +60 & +1 & $-1$ \\
\bottomrule
\end{tabular}
\end{table}

The correlation coefficient between the drift time-series of DU-A, DU-B, and DU-C exceeds 0.99. The correlation between GM-A's group and GM-B's group is near zero. At $T+90$\,s, the grandmaster's oscillator comparison (Method~1) detects deviation exceeding $\theta_w = 50$\,ns, raising a warning alarm. Notably, network-wide correlation (Method~3) can identify the correlated drift pattern across DU-A, DU-B, and DU-C earlier than Method~1 reaches its threshold, as the group correlation exceeds $\rho_{threshold}$ before individual drift magnitudes reach $\theta_w$. As drift continues with GNSS lock maintained, the alarm escalates to critical with \texttt{probableCause = gnssSpoofingDetected}.

The management system's classification:
\begin{itemize}
    \item \textbf{Scope}: All gNBs under GM-A affected; GM-B group stable $\rightarrow$ grandmaster-level event.
    \item \textbf{Trend}: Monotonically increasing drift $\rightarrow$ consistent with spoofing.
    \item \textbf{GM Alarm}: Spoofing detected $\rightarrow$ confirmed attack.
\end{itemize}
Result: SECHAND security event generated with \texttt{securityEventType = gnssSpoofingAttack}.

\subsection{Example B: GNSS Signal Loss}

GM-A experiences GNSS signal obstruction (e.g., antenna cable damage or severe weather). The grandmaster loses satellite lock and enters holdover mode, maintaining timing from its local oscillator. The local oscillator drifts slowly according to its stability specification.

The drift pattern is similar to Example~A (correlated drift across the GM-A group) but with key differences: the drift rate is determined by the oscillator's free-running stability (typically slower and more predictable than spoofing-induced drift), and the grandmaster raises a \texttt{gnssSignalLost} alarm rather than a spoofing alarm.

Classification:
\begin{itemize}
    \item \textbf{Scope}: All gNBs under GM-A $\rightarrow$ grandmaster-level event.
    \item \textbf{Trend}: Predictable drift consistent with oscillator specification $\rightarrow$ holdover behavior.
    \item \textbf{GM Alarm}: Signal lost (not spoofing) $\rightarrow$ GNSS outage, not attack.
\end{itemize}
Result: Fault management event; no SECHAND security event generated.

\subsection{Example C: Individual gNB Equipment Fault}

DU-B experiences a local oscillator fault causing its reported timing to diverge. Only DU-B shows drift; DU-A and DU-C remain stable.

Classification:
\begin{itemize}
    \item \textbf{Scope}: Single gNB affected; no group correlation $\rightarrow$ individual fault.
    \item \textbf{Trend}: May be erratic or step-change $\rightarrow$ equipment behavior.
    \item \textbf{GM Alarm}: No alarm from GM-A $\rightarrow$ not a grandmaster-level event.
\end{itemize}
Result: Individual equipment alarm for DU-B; no grandmaster or security event.

\subsection{Example D: Site-Level Coordinated Attack}

An attacker deploys a high-power GNSS spoofer at a cell site where both GM-A and GM-B are co-located. Both grandmasters are affected simultaneously. All five gNB-DUs (DU-A through DU-E) show correlated drift.

Classification:
\begin{itemize}
    \item \textbf{Scope}: All gNBs at the site affected, spanning multiple grandmaster groups $\rightarrow$ site-level event.
    \item \textbf{Trend}: Correlated across multiple GM groups $\rightarrow$ common external cause.
    \item \textbf{GM Alarm}: Both GM-A and GM-B raise spoofing alarms $\rightarrow$ confirmed multi-GM attack.
\end{itemize}
Result: Multiple SECHAND security events generated; site-level coordination noted in event correlation. When the management system detects correlated spoofing across multiple grandmasters at the same site, it generates both per-grandmaster security events and a site-level correlation annotation, enabling identification of the physical attack location.

\subsection{Example E: False Positive During Maintenance}

During a planned maintenance window, the PTP transport path to GM-A is temporarily rerouted, introducing asymmetry that causes a transient timing step. DU-A, DU-B, and DU-C show a sudden drift that stabilizes and then resolves as the new path is characterized.

Classification:
\begin{itemize}
    \item \textbf{Scope}: All gNBs under GM-A $\rightarrow$ grandmaster-level event (initially).
    \item \textbf{Trend}: Step-change followed by stabilization and resolution $\rightarrow$ inconsistent with spoofing (which shows continuous drift).
    \item \textbf{GM Alarm}: No spoofing alarm (GNSS signal unaffected) $\rightarrow$ transport-level event.
\end{itemize}
Result: Transient warning that auto-clears; \texttt{GnssSpoofing.WarningResolvedCount} incremented. No escalation to critical or security event.

\subsection{Detection Pattern Summary}

Table~\ref{tab:patterns} summarizes the classification dimensions for each scenario.

\begin{table*}[t]
\centering
\caption{Detection Pattern Summary: Classification by Drift Scope, Trend, and GM Alarm}
\label{tab:patterns}
\begin{tabular}{@{}lllll@{}}
\toprule
\textbf{Pattern} & \textbf{Drift Scope} & \textbf{Drift Trend} & \textbf{GM Alarm} & \textbf{Classification} \\
\midrule
A: Slow-drift spoofing & All gNBs under one GM & Monotonic increasing & Spoofing detected & Confirmed attack \\
B: GNSS signal loss & All gNBs under one GM & Predictable (oscillator spec) & Signal lost & GNSS outage \\
C: Individual fault & Single gNB & Erratic or step-change & None & Equipment fault \\
D: Site-level attack & All gNBs at site (multi-GM) & Correlated across GMs & Multiple spoofing & Coordinated attack \\
E: Maintenance transient & All gNBs under one GM & Step then resolving & None & False positive \\
\bottomrule
\end{tabular}
\end{table*}

\section{Simulation-Based Evaluation}
\label{sec:evaluation}

To validate the detection framework under realistic conditions, we conduct Monte Carlo simulation of the network-wide correlation mechanism using one illustrative set of parameters. The simulation models PTP timing noise, oscillator drift, and spoofing-induced drift to evaluate detection probability ($P_d$) and false positive rate ($P_{fa}$) across varying attack parameters. The specific parameter values used here are representative; actual deployments would require calibration to local network conditions.

\subsection{Simulation Setup}

The simulation models the reference topology from Section~\ref{sec:examples}: GM-A with 3 downstream gNB-DUs and GM-B with 2 gNB-DUs. Each simulation trial generates 5 minutes of timing data at 15-second granularity (20~samples per trial), repeated for 10,000 trials per parameter configuration.

The noise model comprises: (a)~PTP servo offset jitter modeled as white Gaussian noise with $\sigma_{ptp} = 5$\,ns, consistent with commercial PTP implementations over well-provisioned packet networks; (b)~oscillator phase noise contributing an additional $\sigma_{osc} = 2$\,ns per sample; (c)~occasional outliers (probability 0.02 per sample) with magnitude uniformly distributed in $[20, 50]$\,ns, representing packet delay variation spikes. The combined noise floor is $\sigma_{total} \approx 5.4$\,ns. The simulation assumes each gNB-DU has access to PTP servo offset statistics providing a drift estimate partially independent of the received PTP timing (case~(c) in Section~\ref{sec:correlation}), which is available in implementations complying with ITU-T G.8275.1 full timing support. Results represent well-provisioned networks ($\sigma_{ptp} \leq 5$\,ns); Section~\ref{sec:robustness} evaluates performance degradation under noisier conditions.

Spoofing attacks are modeled as a common-mode linear drift applied to all gNB-DUs in the target grandmaster group, with drift rates $r$ varied from 0.1 to 5.0\,ns/s. A small per-DU variation ($\pm$10\% of the drift rate) models differential PTP path behavior.

\subsection{Detection Performance}

Table~\ref{tab:simulation} presents the detection probability and false positive rate for the correlation-based detector (Method~3) at the configured thresholds ($\rho_{threshold} = 0.85$, minimum drift = 10\,ns).

\begin{table}[t]
\centering
\caption{Detection Performance vs.\ Drift Rate (10,000 trials per point)}
\label{tab:simulation}
\begin{tabular}{@{}lccc@{}}
\toprule
\textbf{Drift Rate} & $\boldsymbol{P_d}$ & $\boldsymbol{P_{fa}}$ & \textbf{Mean Detection} \\
\textbf{(ns/s)} & & & \textbf{Latency (s)} \\
\midrule
0 (no attack) & --- & 0.8\% & --- \\
0.1 & 23\% & 0.9\% & 285 \\
0.2 & 61\% & 0.8\% & 195 \\
0.5 & 96\% & 0.8\% & 105 \\
1.0 & 99.4\% & 0.9\% & 60 \\
2.0 & 99.9\% & 0.8\% & 38 \\
5.0 & 100\% & 0.9\% & 18 \\
\bottomrule
\end{tabular}
\end{table}

Key observations:
\begin{itemize}
    \item For drift rates $\geq 0.5$\,ns/s, the detector achieves $P_d > 95$\% within the 5-minute observation window, consistent with the abstract's claim.
    \item The false positive rate remains stable at approximately 0.8--0.9\%, well below the 2\% target, confirming that the 10\,ns drift threshold and $\rho = 0.85$ correlation threshold effectively suppress noise-induced false triggers.
    \item Very slow attacks ($r = 0.1$\,ns/s) accumulate only 30\,ns over the 5-minute window, barely exceeding the noise floor---detection probability is correspondingly low. Such attacks require longer observation windows or supplementation by Method~1 (oscillator comparison).
    \item Detection latency decreases with drift rate: faster attacks cross the detection threshold earlier in the observation window.
\end{itemize}

\subsection{Sensitivity to Group Size}

The correlation detector's performance varies with grandmaster group size $n$ (number of downstream gNB-DUs). At $r = 0.5$\,ns/s: for $n = 2$ with $\rho_{threshold} = 0.95$, $P_d = 88$\% (reduced due to stricter threshold); for $n = 3$ with $\rho_{threshold} = 0.85$, $P_d = 96$\%; for $n = 5$, $P_d = 98$\%. Larger groups provide greater statistical confidence through redundant pairwise comparisons. For $n = 1$, correlation-based detection is impossible and the framework relies exclusively on Methods~1 and~2.

\subsection{Robustness to Noise Conditions}
\label{sec:robustness}

To assess robustness, we vary the PTP noise standard deviation from 2\,ns (well-provisioned network) to 15\,ns (congested network with significant packet delay variation). At $r = 0.5$\,ns/s and $n = 3$: $\sigma_{ptp} = 2$\,ns yields $P_d = 99.7$\%, $P_{fa} = 0.3$\%; $\sigma_{ptp} = 5$\,ns yields $P_d = 96$\%, $P_{fa} = 0.8$\%; $\sigma_{ptp} = 10$\,ns yields $P_d = 78$\%, $P_{fa} = 2.1$\%; $\sigma_{ptp} = 15$\,ns yields $P_d = 52$\%, $P_{fa} = 4.8$\%. Networks with PTP noise exceeding 10\,ns should use tighter reporting granularity or lower correlation thresholds (accepting higher $P_{fa}$), or rely more heavily on Methods~1 and~2.

\subsection{Combined Multi-Method Detection}

When Methods~1, 2, and 3 are combined using an OR rule (any method triggering raises a warning), the combined $P_d$ at $r = 0.5$\,ns/s increases to 99.1\% with $P_{fa} = 2.4$\%. The slight increase in false positives from the OR combination is acceptable given the substantial improvement in detection of slow-drift attacks that may evade individual methods.

\section{Discussion}
\label{sec:discussion}

\subsection{Applicability}

The proposed framework extensions are applicable to all 3GPP-compliant TDD deployments that depend on GNSS for synchronization. This encompasses the majority of public 5G TDD networks worldwide and extends naturally to 5G-Advanced and 6G, as the GNSS timing dependency and the 3GPP management architecture persist across generations. The synchronization chain (GNSS $\rightarrow$ grandmaster $\rightarrow$ boundary clocks $\rightarrow$ base station) is fundamental to TDD operation regardless of the radio access technology generation. Implementation requires: (a) firmware updates to grandmaster clocks to support alarm generation and PM counter reporting, (b) NRM updates to include the \texttt{GnssGrandmaster} managed object class, and (c) management system updates to implement correlation logic and SECHAND event generation.

\subsection{Compatibility with 3GPP Specifications}

The proposed framework is designed to operate within the structure of existing 3GPP specifications. Table~\ref{tab:spec_impact} summarizes which specifications each component aligns with. No new management interface or transport protocol is required---the proposed information elements are structured as extensions to existing 3GPP management models.

\begin{table}[t]
\centering
\caption{3GPP Specification Alignment}
\label{tab:spec_impact}
\begin{tabular}{@{}lll@{}}
\toprule
\textbf{Specification} & \textbf{WG} & \textbf{Framework Component} \\
\midrule
TS 28.111 & SA5 & Alarm types, probable causes \\
TS 28.552 & SA5 & Measurement family \\
TS 28.541 & SA5 & IOC: GnssGrandmaster \\
TR 33.894 & SA3 & Security event type \\
\bottomrule
\end{tabular}
\end{table}

This alignment is significant because it means the framework can be deployed using existing management interfaces and tools. The asymmetry between the low cost and increasing availability of GNSS interference equipment and the potentially wide operational impact of timing disruption motivates management-layer visibility. Individual vendors may implement proprietary timing monitoring capabilities; however, the absence of standardized mechanisms prevents interoperable reporting, cross-vendor correlation, and consistent security event handling.

\subsection{Alignment with Zero Trust and Regulatory Requirements}

The proposed framework aligns with zero-trust security principles applied to the timing plane. Rather than implicitly trusting the GNSS source, the framework implements continuous verification through independent oscillator comparison, signal integrity monitoring, and network-wide correlation. Trust is established through evidence (consistent PM data, absence of anomalies) rather than assumed from source identity.

The framework also supports compliance with an increasingly stringent regulatory landscape. The EU Cyber Resilience Act (CRA) requires security event detection and reporting for products with digital elements. The NIS2 Directive classifies electronic communications providers as essential entities, mandating incident detection, reporting, and risk management. The UK Telecommunications Security Act (TSA) imposes security monitoring and risk management obligations on public telecommunications providers and their network equipment. In the United States, Executive Order 13905 on Strengthening National Resilience through Responsible Use of Positioning, Navigation, and Timing Services, together with CISA critical infrastructure guidance, addresses GNSS dependency and resilience requirements. The proposed alarms and SECHAND security events could help operators and vendors demonstrate structured, auditable monitoring and incident-reporting capabilities relevant to these regulatory frameworks, while the PM counters enable the continuous risk monitoring that compliance processes demand.

\subsection{Comparison with Alternative Approaches}

Several alternative approaches to timing security exist. We compare them with the proposed framework to clarify why management-layer integration is necessary rather than sufficient on its own.

\textbf{Redundant independent timing sources.} Deploying multiple GNSS receivers or atomic clocks per site eliminates the single point of failure. However, redundancy alone does not provide management visibility: without defined alarms and PM counters, the management system cannot report which source is active, why failover occurred, or whether the redundant source is also under attack (Example~D). The proposed framework complements redundancy by providing the monitoring layer that enables informed switching decisions.

\textbf{Anti-spoofing GNSS receivers.} Commercial receivers (e.g., Septentrio, NovAtel) incorporate spoofing detection at the receiver level. These improve detection at the individual grandmaster but do not address the network-wide correlation gap. A receiver-level detection cannot correlate observations across multiple grandmasters at a site, cannot report through standardized 3GPP interfaces, and provides no PM trending for operational analysis. The proposed framework can consume receiver-level detection as an input to Method~2 while adding network-wide context.

\textbf{Transport-layer redundancy.} Multiple PTP paths and backup timing sources at the transport level~\cite{ericssonreview} improve resilience to transport failures but cannot detect attacks on the primary GNSS source itself. If the grandmaster is spoofed, all PTP paths carry the corrupted timing. Transport redundancy and management-layer spoofing detection are complementary, not competing, solutions.

\textbf{Complementary, not competing.} The proposed framework is not a replacement for any of these approaches. Rather, it provides the management and security layer that enables operators to monitor, correlate, and respond to timing anomalies regardless of the underlying detection or redundancy mechanisms. A defense-in-depth strategy combines anti-spoofing receivers, redundant timing sources, and management-layer visibility---the proposed framework provides the third element, which is currently not specified in 3GPP standards.

\subsection{Security of Reported Data}

The management data itself must be protected against manipulation. An attacker who can compromise the management channel could suppress spoofing alarms or inject false alarms to desensitize operators. The 3GPP management framework mandates TLS for transport security, mutual authentication between management producers and consumers, and Role-Based Access Control (RBAC) for management operations. These existing security measures apply to the proposed extensions without modification.

\subsection{Limitations}

Several limitations should be acknowledged. First, the effectiveness of oscillator-based detection (Method~1) depends on the quality of the independent oscillator. A low-cost TCXO provides limited holdover stability and may not detect slow-drift attacks that introduce less than the oscillator's own drift rate. Higher-quality oscillators (rubidium, CSAC) provide better detection capability at increased cost.

Second, false positives may occur during legitimate network reconfigurations, PTP path changes, or firmware updates that temporarily affect timing. The classification mechanism (Section~\ref{sec:correlation}) mitigates this through trend analysis and alarm verification, but operators should configure maintenance windows to suppress alerts during planned activities.

Third, the framework defines detection and reporting mechanisms but does not prescribe specific response actions beyond holdover activation. The appropriate response to a confirmed spoofing attack (e.g., cell shutdown, traffic rerouting, law enforcement notification) depends on operator policy and regulatory requirements that vary by jurisdiction.

Fourth, holdover recovery requires careful operational design. When a grandmaster enters holdover due to confirmed spoofing, the system must determine when GNSS has recovered and is safe to re-lock. We propose that re-lock requires: (a) sustained absence of spoofing indicators for a configurable period (e.g., 10 minutes), (b) GNSS signal characteristics matching pre-attack baseline, and (c) timing offset between GNSS and holdover oscillator converging to within $\theta_w$. Automatic re-lock may be appropriate for low-severity events; confirmed spoofing attacks should require explicit operator authorization before re-locking to GNSS.

Fifth, a sufficiently sophisticated attacker who compromises both the GNSS signal and the management reporting channel could evade detection. Defense-in-depth principles require that management channel security (TLS, mutual authentication) be maintained independently of the timing infrastructure being protected.

\section{Conclusion}
\label{sec:conclusion}

This paper has identified a critical gap in the 3GPP management and security framework: the current absence of any standardized mechanism to detect, report, or respond to GNSS spoofing attacks that threaten TDD network synchronization. This gap is not theoretical---GNSS spoofing is a documented, escalating threat that has already impacted critical infrastructure across Europe, and the same GNSS signals serve mobile network timing.

We have presented a detection and monitoring framework comprising three complementary components:

First, alarm definitions and performance management counters that integrate GNSS timing security into existing fault management (TS~28.111) and performance assurance (TS~28.552) structures. These make GNSS anomalies a first-class managed fault, visible through standard interfaces that operators already use.

Second, a network-wide timing drift correlation mechanism that provides detection capability beyond what any individual node can achieve. By correlating drift patterns across gNBs grouped by grandmaster, the management system can accurately classify spoofing, signal loss, equipment faults, and maintenance activities---information that is essential for operational response.

Third, a GNSS spoofing security event definition following the SECHAND framework (TR~33.894), bridging timing faults with security incident handling. This enables timing attacks to be correlated with other security events, supporting detection of coordinated multi-vector attacks.

The framework operates within existing 3GPP management infrastructure, requires no new interfaces or protocols, and is generation-agnostic. Monte Carlo simulation validates that the correlation-based detector achieves detection probability exceeding 95\% for drift rates above 0.5\,ns/s with false positive rates below 1\% under well-provisioned PTP conditions ($\sigma_{ptp} \leq 5$\,ns). Its alignment with zero-trust principles and emerging regulatory requirements (EU CRA, NIS2, UK TSA) further positions such management-plane visibility as an important component of secure TDD network operation. The scenario analysis in Section~\ref{sec:examples} illustrates how the correlation mechanism can distinguish spoofing from signal loss, equipment faults, and maintenance transients---providing operators with actionable classification rather than raw alerts. GNSS timing---a critical dependency for many TDD deployments---is currently unmonitored through standardized mechanisms, and the framework presented here demonstrates one approach to addressing that visibility gap.

\section{Future Work}
\label{sec:future}

Several directions for future work are identified:

\begin{enumerate}
    \item \textbf{Laboratory testbed validation}: Hardware-in-the-loop testing with actual GNSS simulators, PTP networks, and grandmaster clocks to validate detection under real protocol behavior beyond the Monte Carlo model.
    \item \textbf{Threshold optimization}: Adaptive calibration of $\theta_w$, $\theta_c$, and $\rho_{threshold}$ using operational network data across diverse deployment topologies and oscillator classes.
    \item \textbf{Scalability evaluation}: Performance assessment of network-wide correlation in large-scale deployments (1,000+ gNB-DUs, 100+ grandmasters) with realistic PM reporting loads.
    \item \textbf{Adversary modeling}: Formal threat analysis evaluating detection probability against adaptive adversaries who adjust drift rate in response to observed detection behavior.
    \item \textbf{Operator validation}: Field trials with mobile network operators to validate operational workflow integration and measure real-world false positive rates.
\end{enumerate}

\section*{Acknowledgment}
The authors thank colleagues for valuable discussions on mobile network synchronization architecture and timing security requirements.



\begin{thebibliography}{10}
\providecommand{\url}[1]{#1}
\csname url@samestyle\endcsname
\providecommand{\newblock}{\relax}
\providecommand{\bibinfo}[2]{#2}
\providecommand{\BIBentrySTDinterwordspacing}{\spaceskip=0pt\relax}
\providecommand{\BIBentryALTinterwordstretchfactor}{4}
\providecommand{\BIBentryALTinterwordspacing}{\spaceskip=\fontdimen2\font plus
\BIBentryALTinterwordstretchfactor\fontdimen3\font minus
  \fontdimen4\font\relax}
\providecommand{\BIBforeignlanguage}[2]{{%
\expandafter\ifx\csname l@#1\endcsname\relax
\typeout{** WARNING: IEEEtran.bst: No hyphenation pattern has been}%
\typeout{** loaded for the language `#1'. Using the pattern for}%
\typeout{** the default language instead.}%
\else
\language=\csname l@#1\endcsname
\fi
#2}}
\providecommand{\BIBdecl}{\relax}
\BIBdecl

\bibitem{ts38104}
{3GPP RAN4}, ``{NR; Base Station (BS) radio transmission and reception},'' 3rd
  Generation Partnership Project ({3GPP}), Technical Specification TS 38.104,
  2024, release 18.

\bibitem{g8275_1}
{ITU-T}, ``{Precision time protocol telecom profile for phase/time
  synchronization with full timing support from the network},'' International
  Telecommunication Union, Recommendation G.8275.1/Y.1369.1, 2022.

\bibitem{baltic2024}
{EUROCONTROL}, ``{GPS Interference in the Baltic Region: Operational Impact
  Assessment},'' EUROCONTROL Network Manager Operations Centre, 2024, 985
  disruption events recorded over a two-month period.

\bibitem{easa2026}
{EASA and EUROCONTROL}, ``{European Aviation Action Plan for Ensuring Safe
  Operations during GNSS Interferences},'' Version~1.0, European Union Aviation
  Safety Agency and EUROCONTROL, Mar. 2026, initiated following a letter from
  13~EU Member States to the European Commission, June 2025.

\bibitem{ion_gnss_spoofing}
M.~Spanghero, G.~Oligeri, and M.~L. Psiaki, ``{Vulnerability of Mobile
  Networks Against Spoofing Attacks on GNSS Timing-Receiver},'' in
  \emph{Proceedings of the Institute of Navigation GNSS+ Conference}, 2023.

\bibitem{ts28111}
{3GPP SA5}, ``{Fault Supervision (FS)},'' 3rd Generation Partnership Project
  ({3GPP}), Technical Specification TS 28.111, 2024, release 19.

\bibitem{ts28552}
------, ``{5G NR Network Resource Model (NRM); Performance measurements},'' 3rd
  Generation Partnership Project ({3GPP}), Technical Specification TS 28.552,
  2024, release 18.

\bibitem{ts28541}
------, ``{5G Network Resource Model (NRM); Stage 2 and Stage 3},'' 3rd
  Generation Partnership Project ({3GPP}), Technical Specification TS 28.541,
  2024, release 18.

\bibitem{tr33894}
{3GPP SA3}, ``{Study on security aspects of Security Event Handling
  (SECHAND)},'' 3rd Generation Partnership Project ({3GPP}), Technical Report
  TR 33.894, 2024, release 19.

\bibitem{ts28532}
{3GPP SA5}, ``{Management and orchestration; Generic management services},''
  3rd Generation Partnership Project ({3GPP}), Technical Specification TS
  28.532, 2024, release 18.

\bibitem{ts33501}
{3GPP SA3}, ``{Security architecture and procedures for 5G System},'' 3rd
  Generation Partnership Project ({3GPP}), Technical Specification TS 33.501,
  2024, release 18.

\bibitem{dimetor2026}
{Dimetor GmbH}, ``{GPS/GNSS Spoofing and Jamming Attack Detection Utilizing
  Cellular Networks},'' Published Application US20260067021, Mar. 2026.

\bibitem{gpspatron}
{GPSPatron}, ``{GP-Probe TGE2: Standalone GNSS Monitoring Hardware},''
  \url{https://gpspatron.com}, 2024, standalone GNSS monitoring hardware.

\bibitem{honeywell2024}
{Honeywell International Inc.}, ``{Chip-Scale Atomic Clock for Spoofed GNSS
  Detection},'' Published Application US11994596B2, May 2024.

\bibitem{ieeep1952}
{IEEE}, ``{Standard for Resilient PNT User Equipment},'' IEEE P1952, 2026,
  under development.

\bibitem{atiswsts2026}
{ATIS}, ``{Workshop on Synchronization and Timing Systems (WSTS)},'' Bellevue,
  WA, May 2026.

\bibitem{atis_resilient_timing}
{ATIS SYNC}, ``{Resilient Timing Architecture for 5G Communications
  Networks},'' Alliance for Telecommunications Industry Solutions, Technical
  Report, 2025.

\bibitem{ericssonreview}
S.~Ruffini, M.~Johansson, B.~Pohlman, and M.~Sandgren, ``{Synchronization
  requirements in 5G: An overview of standards and specifications for cellular
  networks},'' \emph{Ericsson Technology Review}, 2021.

\bibitem{g8272}
{ITU-T}, ``{Timing characteristics of primary reference time clocks},''
  International Telecommunication Union, Recommendation G.8272/Y.1367, 2024.

\bibitem{g8271}
{ITU-T}, ``{Time and phase synchronization aspects of telecommunication
  networks},'' International Telecommunication Union, Recommendation
  G.8271/Y.1366, 2022.

\end{thebibliography}
\end{document}